\documentclass{article}
\usepackage{fullpage}
\usepackage{authblk}
\usepackage{amsmath}
\usepackage{amssymb}
\usepackage{graphicx}
\usepackage[colorlinks=true]{hyperref}
\usepackage[font=small,labelfont=bf]{caption}

\begin{document}

\title{Dispersivity calculation in digital twins of multiscale porous materials using the micro-continuum approach}

\author[1]{Julien Maes}
\author[1]{Hannah P. Menke}
\affil[1]{Institute of GeoEnergy Engineering, Heriot-Watt University, Edinburgh, U.K.}

\maketitle

\begin{abstract}
The micro-continuum  method is a novel approach to simulate flow and transport in multiscale porous materials. For such materials, the domain can be divided into three sub-domains depending on the local porosity $\varepsilon$: fully resolved solid phase, for which $\varepsilon=0$, fully resolved pores, for which $\varepsilon=1.0$, and unresolved pores, for which $0<\varepsilon<1.0$. For such domains, the flow can be solved using the Darcy-Brinkman-Stokes (DBS) equation, which offers a seamless transition between unresolved pores, where flow is described by Darcy's law, and resolved pores, where flow is described by the Navier-Stokes equations. Species transport can then be modelled using a volume-averaged equation. In this work, we present a derivation of the closure problem for the micro-continuum approach. Effective dispersivity tensors can then be calculated through a multi-stage process. First, high resolution images are chosen for characterizing the structure of the unresolved pores. Porosity, permeability and effective dispersivity for the unresolved part are calculated by solving a closure problem based on Direct Numerical Simulation (DNS) in the high-resolution images. The effective dispersivity is then expressed as a function of the P\'eclet number, which describes the ratio of advective to diffusive transport. This relationship, along with porosity and permeability, is then integrated into the multiscale domain and the effective dispersivity tensor for the full image is calculated. Our novel method is validated by comparison with the numerical solution obtained for a fully-resolved simulation in a multiscale 2D micromodel. It is then applied to obtain an effective dispersivity model in digital twins for two multiscale materials: hierarchical ceramic foams and microporous carbonate rocks. 
\end{abstract}

\section{Introduction}
\label{Sect:Intro}
Understanding flow and transport in porous material is of great importance for a wide range of engineering applications, such as carbon capture and storage \cite{2011-Nordbotten}, geothermal systems \cite{2015-Pandey}, catalytic beds \cite{2016-Das}, electrolyzers \cite{2021-Xu} and fuel cells \cite{2022-Jiao}. Although these applications are controlled by transport phenomena that occurs at the pore-scale \cite{2015-Pak,2017-Menke,2017-Reynolds}, modelling and optimising production scenarios relies on a macroscopic description based on a Representative Elementary Volume (REV) \cite{2010-AlRaoush}. Digital Twin Technology \cite{2013-Andrae} can then be employed to calculate upscaling parameters, such as permeability \cite{2013b-Andrae} and effective dispersivity \cite{2021-Soulaine}, by performing numerical simulation on X-ray computed micro-tomography (micro-CT) images of an REV using Direct Numerical simulation (DNS) or Pore Network Modelling (PNM).

Effective dispersivity is the physical process resulting from the variations in the velocity field which, combined with molecular diffusion, creates spreading of a solute distribution at the macro-scale. Various simulation approaches have been used to calculate effective dispersivity coefficients from micro-CT images of porous materials, including the random walk method \cite{2002-Nakashima},  the time domain random walk method \cite{2002-Delay,2012-Dentz}, the continuous time random walk method \cite{2006-Berkowitz,2006-Bijeljic},  fitting a breakthrough curve \cite{2010-Zaretskiy,2021-OrtegaRamirez} or solving a closure problem based on DNS \cite{1983-Carbonel,2021-Soulaine}. All these methods provide accurate effective dispersivity coefficients as long as the pore surfaces are accurately resolved in the micro-CT image \cite{2013-Richmond,2016-Noetinger}.

However, porous materials are often multiscale, i.e. they have several pore structures (e.g. micro-pores, macro-pores, channels) that have length scales that differ by orders of magnitude. For such materials, obtaining a micro-CT image that is an REV of all structures and the connectivity between them is challenging, if not impossible, because such an image would need to include an REV of the largest pore-size structure but have enough resolution to characterise accurately the smallest of pores. Even the most advanced X-ray apparatus in the world cannot provide such images, and even if they could, the domain would be much too large to run simulations over them. Therefore, multiscale porous materials can only be investigated numerically using a multiscale workflow. 

A Digital Twin for a multiscale porous material is based on a multiscale image, i.e. an image that can be divided into three subdomains: the fully resolved pores, the fully resolved solid and the unresolved pores. In addition, the unresolved pores are described by at least one image at higher resolution \cite{2013-Jiang,2021-Moslemipour,2022-Menke}. If the image is a two-scale image, then the higher-resolution image fully resolves the smallest pores. If the image has more than two scales, then the higher-resolution image is itself a multiscale image. Effective flow and transport properties in a multiscale image can be calculated using a multiscale workflow, where the flow and transport properties of the higher-resolution image are calculated first, and then integrated into the multiscale image. Two categories of such multiscale workflows have been employed in the past, those based on multiscale pore-network models \cite{2021-Ruspini,2021-Moslemipour}, and those based on the micro-continuum method \cite{2024-Soulaine}. 

The micro-continuum method is a novel approach to simulate flow and transport in multiscale porous materials. Within this approach, each voxel in the multiscale image (resolved pores, resolved solid and unresolved pores) is characterised by a set of effective properties (e.g. porosity, permeability, dispersivity). Flow is then solved using the Darcy-Brinkman-Stokes (DBS) equation \cite{1949-Brinkman,2016a-Soulaine}, which offers a seamless transition between unresolved pores, for which flow is described by Darcy's law, and resolved pores, for which flow is described by the Navier-Stokes equations, while the solid phase can be penalized by setting a near-zero permeability. The micro-continuum method has been applied to simulate flow and transport in multiscale images \cite{2022-Menke,2016-Soulaine,2020-Carrillo}. The micro-continuum method can also be applied to single-scale image, an approach that has advantages in term of modelling solid dissolution and precipitation \cite{2017-Soulaine,2021-Yang,2022-Maes} as well as modelling heat transfer between fluid and solid \cite{2021c-Maes}.   

In this paper, we present a derivation of the closure problem for dispersivity with the micro-continuum method which can be applied to multiscale images. The mathematical models are presented in section \ref{section:model}. The derivation of the closure problem and calculation of the effective dispersivity for multiscale domain are presented in section \ref{section:disp}, and their implementation in GeoChemFoam (\href{www.github.com/geochemfoam}{www.github.com/geochemfoam}), our open-source pore-scale reactive transport solver, is presented in section \ref{section:imp}.  Our novel approach is validated by comparison with the effective dispersivity calculated with a closure problem based on DNS for a 2D multiscale micromodel in section \ref{section:valid}. Finally, our novel method is used to obtain an effective dispersivity model for two different types of geometry: hierarchical porous materials representing ceramic foams and a microporous carbonate rock (section \ref{section:app}). 

\section{Mathematical models}
\label{section:model}
\subsection{Governing equations}

We consider a spatially periodic porous medium. The governing equations consider flow, convective transport and molecular diffusion within the pore-space. The domain $\Omega$ is partitioned into fluid $\Omega_f$ and solid $\Omega_s$. We assume that fluid motion in $\Omega_f$ is governed by the steady-state incompressible Navier-Stokes equations
\begin{equation}
\label{Eq:cont}\nabla\cdot\mathbf{u} = 0,
\end{equation}
\begin{equation}
\nabla\left(\mathbf{u}\otimes\mathbf{u}\right)=-\nabla p +\nu\nabla^2\mathbf{u}+\mathbf{S}_u,\label{Equ:momentum}
\end{equation}
with a no-flow no-slip condition at the fluid-solid interface $\Gamma$,
\begin{equation}\label{Equ:bcu}
\mathbf{u}=0 \hspace{0.5cm} \text{at $\Gamma$},
\end{equation}
where $\mathbf{u}$ (m/s) is the velocity, $p$ (m$^2$/s$^2$) is the kinematic pressure, $\nu$ (m$^2$/s) is the kinematic viscosity and $\mathbf{S}_u$ is the external body force. The concentration $c$ (kmol/m$^3$) of a species in the system satisfies an advection-diffusion equation
\begin{equation}\label{Eq:concentration}
\frac{\partial c}{\partial t}+ \nabla \cdot \left( c\mathbf{u} \right) = \nabla\cdot\left(D\nabla c\right),
\end{equation}
where  $D$ (m$^2$/s) is the diffusion coefficient. In the absence of reaction and adsorption, the boundary condition at the fluid-solid interface is given by
\begin{equation}
 \mathbf{n}_s\cdot  \nabla c=0 \hspace{0.5cm} \text{at $\Gamma$},
\end{equation}
where  $\mathbf{n_s}$ is the normal vector to the fluid-solid interface

\subsection{Micro-continuum approach}

In the micro-continuum (MC) approach, the entire domain $\Omega$ is considered, i.e fluid $\Omega_f$ and solid $\Omega_s$, and the fluid-solid interface is described in terms of $V_f$ and $V_s$, the volume of fluid and solid phase in each control volume $V$, and their volume fractions $\varepsilon=V_f/V$ and $\varepsilon_s=1-\varepsilon$. For a quantity $\beta$ defined in $\Omega_f$, we define $\beta_f$, the intrinsic fluid quantity in each control volume as
\begin{equation}
 \beta_f=\frac{1}{V_f}\int_{V_f}\beta dV.
\end{equation}
The averaging process results in the DBS equation \cite{2016a-Soulaine}
\begin{equation}\label{Eq:continuityDBS}
\nabla\cdot\mathbf{u}_D=0
\end{equation}
\begin{equation}\label{Eq:momentumDBS}
 \frac{1}{\varepsilon}\nabla\left(\frac{\mathbf{u}_D\otimes\mathbf{u}_D}{\varepsilon}\right)=-\nabla p_f +\nu\nabla^2{\mathbf{u}_D}-\tilde{\nu} k^{-1}{\mathbf{u}_D}+\mathbf{S}_{uf},
\end{equation}
with a no-slip condition at resolved solid boundaries $\Gamma_f$
\begin{equation}\label{Equ:bcuf}
\mathbf{u}_D=0 \hspace{0.5cm} \text{at $\Gamma_f$},
\end{equation}
where $\mathbf{u}_D=\varepsilon\mathbf{u}_f$ is the Darcy velocity, $\tilde{\nu}$ is the effective viscosity in the unresolved pores and $k$ (m$^2$) is the permeability in the control volume. $\nu k^{-1}{\mathbf{u}_D}$ represents the momentum exchange between the fluid and the solid phase, i.e. the Darcy resistance. This term is dominant in the unresolved pores, but vanishes in the resolved pores. In the solid phase, the permeability is set to a number small enough to obtain a no-slip boundary condition at the fluid-solid interface. Defining the exact form of the effective viscosity is an unresolved research question \cite{2007-ValdesParada,2007-Liu}. In this work, we adopt a simple formulation $\tilde{\nu}=\nu$.

For the species transport, the mass-balance equation averaged over the control volume gives
\begin{equation}\label{Eq:concentrationDBS}
 \frac{\partial \varepsilon_f{c_f}}{\partial t}+\nabla\cdot\left({\varepsilon\mathbf{u}_f}{c_f}\right)-\nabla\cdot\left( \varepsilon D^*\nabla{c_f}\right)=0,
\end{equation}
where $\varepsilon D^*$ (m$^2$/s) is the effective diffusion coefficient. The effective diffusion coefficient takes into account the reduction of the total diffusion due to the presence of solid phase. The boundary condition at resolved solid surfaces $\Gamma_f$ reads
\begin{equation}\label{Eq:bcf}
 \mathbf{n}_s \cdot \nabla c_f=0 \hspace{0.5cm} \text{at $\Gamma_f$}.
\end{equation}

\section{Dispersion calculation}
\label{section:disp}
\subsection{Volume-averaged equation}

We assume that the characteristic length of the averaging volume $r$ is large compared to the pore-sale length $l$, i.e. $l \ll r$. We then define the volume-averaging operator $\langle\cdot\rangle^f$ of a quantity $\beta$ defined in $\Omega_f$ as
\begin{equation}
 \langle\beta\rangle^f=\frac{1}{V_{\Omega_f}}\int_{\Omega_f}\beta dV.
\end{equation}
In addition, we define the volume-averaging operator $\langle\cdot\rangle$ of a quantity $\theta$ defined in $\Omega$ as
\begin{equation}
 \langle\theta\rangle=\frac{1}{V_{\Omega}}\int_{\Omega}\theta dV.
\end{equation}
It can be shown that
\begin{equation}\label{Eq:averagef}
 \langle\varepsilon\beta_f\rangle=\phi\langle\beta\rangle^f,
\end{equation}
where $\phi=\langle\varepsilon\rangle$ is the total porosity of the averaged domain. Using the spatial averaging theorem
\begin{equation}
\langle \nabla\cdot\theta\rangle=\nabla\cdot\langle\theta\rangle+\frac{1}{V_\Omega}\int_{\Gamma_f}\theta\cdot\mathbf{n}_sdA,
\end{equation}
the volume-averaging operator is applied to the continuity equation (Eq. (\ref{Eq:continuityDBS})), yielding
\begin{equation}
\phi\nabla\cdot\langle\mathbf{u}\rangle^f=0.
\end{equation}
Applying the volume-averaging operator to the volume-averaged species concentration equation (Eq. (\ref{Eq:concentrationDBS})), we obtain
\begin{equation}\label{Eq:volumeAveraged1}
\frac{\partial\langle\varepsilon c_f\rangle}{\partial t}+\nabla\cdot\langle\varepsilon\mathbf{u}_fc_f\rangle=\nabla\cdot\langle \varepsilon D^*\nabla c_f\rangle.
\end{equation}
In Eq. (\ref{Eq:volumeAveraged1}), we used that $\mathbf{u}_f$ and $D\nabla c_f$ are zero on resolved solid boundaries. In order to simplify each term in the equation, we introduced the spatial deviations:
\begin{equation}\label{Eq:ctilde}
\tilde{c}_f=c_f - \langle c\rangle^f
\end{equation}
\begin{equation}\label{Eq:utilde}
\tilde{\mathbf{u}}_f=\mathbf{u}_f-\langle\mathbf{u}\rangle^f 
\end{equation}
\begin{equation}\label{Eq:Dtilde}
\tilde{D}_f=D^*-\frac{\langle \varepsilon D^* \rangle}{\phi} 
\end{equation}
Using Eq. (\ref{Eq:ctilde}) and (\ref{Eq:utilde}) in the advection term of Eq. (\ref{Eq:volumeAveraged1})
\begin{align}
\nabla\cdot\langle\varepsilon\mathbf{u}_fc_f\rangle&=\nabla\cdot\langle\varepsilon\langle\mathbf{u}\rangle^f\langle c\rangle^f\rangle+\nabla\cdot\langle\varepsilon\tilde{\mathbf{u}}_f\langle c\rangle^f\rangle+\nabla\cdot\langle\langle\mathbf{u}\rangle^f\varepsilon\tilde{c}_f\rangle+\nabla\cdot\langle\varepsilon\tilde{\mathbf{u}}_f\tilde{c}_f\rangle\\
&=\phi\langle\mathbf{u}\rangle^f\cdot\nabla\langle c\rangle^f+\nabla\cdot\langle\varepsilon\tilde{\mathbf{u}}_f\tilde{c}_f\rangle \label{Eq:averageUc}
\end{align}
In Eq. (\ref{Eq:averageUc}), we used that the average of an already averaged quantity is the same quantity, and that the averages of $\varepsilon\tilde{c}_f$ and $\varepsilon\tilde{\mathbf{u}}_f$ are zero. This is correct provided that the averaged quantities vary over a characteristic length $L$ that is much greater than the size of the averaging volume $r$, which is itself much greater than the pore characteristic length $l$,
\begin{equation}\label{Eq:big}
L\gg r\gg l.
\end{equation}
Similarly, the diffusion term can be written as
\begin{align}
\nabla\cdot\langle \varepsilon D^* \nabla c_f\rangle&=\nabla\cdot\langle \varepsilon D^*\nabla\langle c\rangle^f\rangle+\nabla\cdot\langle \varepsilon D^*\nabla\tilde{c}_f\rangle\\
&=\langle \varepsilon D^*\rangle\nabla^2\langle c\rangle^f + \nabla\cdot\langle \varepsilon D^*\nabla\tilde{c}_f\rangle \label{Eq:averageDiff}. 
\end{align}
To obtain Eq. (\ref{Eq:averageDiff}), we assumed that the domain is a REV for porosity, so that $\nabla\phi$=0. Therefore, Eq. (\ref{Eq:volumeAveraged1}) can be recast into
\begin{equation}\label{Eq:volumeAveraged2}
\phi\frac{\partial\langle c\rangle^f}{\partial t}+\phi\langle\mathbf{u}\rangle^f\cdot\nabla\cdot\langle c\rangle^f=\langle \varepsilon D^*\rangle\nabla^2\langle c\rangle^f + \nabla\cdot\langle \varepsilon D^*\nabla\tilde{c}_f\rangle-\nabla\cdot\langle\varepsilon\tilde{\mathbf{u}}_f\tilde{c}_f\rangle
\end{equation}
Eq. (\ref{Eq:volumeAveraged2}) is the governing equation for the average concentration of species $c$ in the fluid. The three terms on the right side correspond to the contribution of molecular diffusion, tortuosity and hydrodynamics. Since the governing equation depends on $\tilde{c}_f$, closure is required to calculate the dispersivity tensor.

\subsection{Closure problem}

To obtain the governing equation for the spatial variation $\tilde{c}_f$, we multiply Eq. (\ref{Eq:volumeAveraged2}) by $\varepsilon$, divide it by $\phi$, subtract it from Eq. (\ref{Eq:concentrationDBS}) and use the definition of spatial deviations (Eq. (\ref{Eq:ctilde}), (\ref{Eq:utilde})) and (\ref{Eq:Dtilde})). This leads to
\begin{equation}\label{Eq:closure1}
\begin{aligned}
 \varepsilon\frac{\partial \tilde{c}_f}{\partial t}+ \varepsilon\mathbf{u}_f\cdot\nabla \tilde{c}_f+ \varepsilon\tilde{\mathbf{u}}_f\cdot\nabla \langle c\rangle^f&=\nabla\cdot\left(\varepsilon D^* \nabla\tilde{c}_f\right)+\nabla\left(\varepsilon D^*\right)\cdot\nabla\langle c \rangle^f \\
 & +\varepsilon\tilde{D}_f\nabla^2\langle c\rangle^f - \frac{\varepsilon}{\phi}\nabla\cdot\langle \varepsilon D^*\nabla\tilde{c}_f\rangle+\frac{\varepsilon}{\phi}\nabla\cdot\langle\varepsilon\tilde{\mathbf{u}}_f\tilde{c}_f\rangle.
\end{aligned}
\end{equation} 
Using the comparison of the characteristic length (Eq. (\ref{Eq:big})), we note that
\begin{equation}
\nabla\cdot\langle\varepsilon\tilde{\mathbf{u}}_f\tilde{c}_f\rangle\ll \varepsilon \mathbf{u}_f\cdot\nabla \tilde{c}_f,\\
\end{equation}
\begin{equation}
\nabla\cdot\langle \varepsilon D^*\nabla\tilde{c}_f\rangle \ll \nabla\cdot\left(\varepsilon D^*\nabla\tilde{c}_f\right),\\
\end{equation}
\begin{equation}
\varepsilon\tilde{D}_f\nabla^2\langle c\rangle^f \ll \nabla\cdot\left(\varepsilon D^* \nabla\tilde{c}_f\right),\\
\end{equation}
so the last three terms in Eq. (\ref{Eq:closure1}) can be neglected \cite{1998-Quintard}. This leads to
\begin{equation}\label{Eq:closure2}
 \varepsilon\frac{\partial \tilde{c}_f}{\partial t}+\varepsilon\mathbf{u}_f\cdot\nabla \tilde{c}_f+ \varepsilon\tilde{\mathbf{u}}_f\cdot\nabla \langle c\rangle^f=\nabla\cdot\left(\varepsilon D^*\nabla\tilde{c}_f\right)+\nabla\left(\varepsilon D^*\right)\cdot\nabla\langle c \rangle^f ,
\end{equation}
with the boundary conditions at resolved solid surfaces
\begin{equation}
\mathbf{n}_s \cdot \nabla \tilde{c}_f=-\mathbf{n}_s \cdot \nabla\langle c \rangle^f  \hspace{0.5cm} \text{at $\Gamma_f$}.
\end{equation}
A solution of this equation can be expressed in the form \cite{1983-Carbonel,1998-Quintard}
\begin{equation}\label{Eq:closurevar}
\tilde{c}_f=\mathbf{f}\cdot\nabla\langle c \rangle^f,
\end{equation}
where $\mathbf{f}$ (m) is called the closure variable. Using the separation of scales (Eq. (\ref{Eq:big})), we note that
\begin{equation}\label{Eq:nablaf}
\nabla\tilde{c}_f=\nabla \mathbf{f}\cdot\nabla\langle c \rangle^f.
\end{equation}
Finally, we assume that the characteristic time-scale of variation of $\tilde{c}_f$ is small compared to  the characteristic time-scale of variation of $\langle c\rangle^f$. This can be shown by noting that the operators are the same (advection and diffusion), but applied on a small characteristic length for $\tilde{c}_f$ ($l$) than for $\langle c\rangle^f$ (L) \cite{2003-Wood}. Therefore,
\begin{equation}
\mathbf{f}\cdot\frac{\partial \nabla\langle c \rangle^f}{\partial t}\ll\frac{\partial \mathbf{f}}{\partial t}\cdot\nabla\langle c \rangle^f,
\end{equation}
and the closure variable satisfies
\begin{equation}\label{Eq:closureT}
\varepsilon\frac{\partial \mathbf{f}}{\partial t}+\varepsilon\mathbf{u}_f\cdot\nabla \mathbf{f}+ \varepsilon\tilde{\mathbf{u}}_f=\nabla\cdot\left(\varepsilon D^* \nabla \mathbf{f}\right)+\nabla\left(\varepsilon D^*\right),
\end{equation}
with the boundary conditions at resolved solid surfaces
\begin{equation}
 \mathbf{n}_s\cdot\nabla \mathbf{f}=-\mathbf{n}_s \hspace{0.5cm} \text{at $\Gamma_f$}.
\end{equation}
In this paper, we will only calculate steady-state effective dispersivity coefficient. In this case, the closure problem can be solved in its steady-state, i.e.
\begin{equation}\label{Eq:closureFinal}
\varepsilon\mathbf{u}_f\cdot\nabla \mathbf{f}+ \varepsilon\tilde{\mathbf{u}}_f=\nabla\cdot\left(\varepsilon D^* \nabla \mathbf{f}\right)+\nabla\left(\varepsilon D^*\right).
\end{equation} 
We note that in the case where the image is fully resolved, $\varepsilon=1.0$, $D^*=D$, and the closure problem is identical to the one presented in \cite{2021-Soulaine}.

\subsection{Dispersion tensor}
The closed form of the macroscopic transport equation can be obtained by substituting Eq. (\ref{Eq:closurevar}) in Eq. (\ref{Eq:volumeAveraged2}) 
\begin{equation}\label{Eq:closedTransp}
\phi\frac{\partial\langle c\rangle^f}{\partial t}+\phi\langle\mathbf{u}\rangle^f\cdot\nabla\cdot\langle c\rangle^f=\langle \varepsilon D^*\rangle\nabla^2\langle c\rangle^f + \nabla\cdot\langle \varepsilon D^*\nabla \mathbf{f}\cdot\nabla\langle c \rangle^f\rangle-\nabla\cdot\langle\varepsilon\tilde{\mathbf{u}}_f\mathbf{f}\cdot\nabla\langle c \rangle^f\rangle
\end{equation}
Due to the separation of scales constraint (Eq. (\ref{Eq:big})), $\nabla\langle c \rangle^f$ can be treated as a constant within the averaging volume. Therefore
\begin{equation}\label{Eq:closedTransp2}
\phi\frac{\partial\langle c\rangle^f}{\partial t}+\phi\langle\mathbf{u}\rangle^f\cdot\nabla\cdot\langle c\rangle^f=\langle \varepsilon D^*\rangle\nabla^2\langle c\rangle^f + \nabla\cdot\left(\langle \varepsilon D^*\nabla \mathbf{f}\rangle\cdot\nabla\langle c \rangle^f\right)-\nabla\cdot\left(\langle\varepsilon\tilde{\mathbf{u}}_f\mathbf{f}\rangle\cdot\nabla\langle c \rangle^f\right).
\end{equation}
It is convenient to define the average diffusion tensor
\begin{equation}
\underline{\mathcal{D}}=\frac{\langle \varepsilon D^*\rangle\underline{\mathcal{I}}}{\phi},
\end{equation}
where $\underline{\mathcal{I}}$ is the identity tensor, the tortuosity tensor
\begin{equation}
\underline{\boldsymbol\tau}=\frac{\langle \varepsilon D^*\nabla \mathbf{f}\rangle}{\phi},
\end{equation}
and the hydrodynamic dispersion tensor
\begin{equation}
\underline{\mathbf{D}}=-\frac{\langle\varepsilon\tilde{\mathbf{u}}_f\mathbf{f}\rangle}{\phi}.
\end{equation}
Finally, assuming that the domain is an REV for the tortuosity and hydrodynamic dispersion tensors, i.e. $\nabla\cdot\underline{\boldsymbol\tau}=0$ and $\nabla\cdot\underline{\mathbf{D}}=0$, the final form of the macroscopic transport equation is
\begin{equation}\label{Eq:closedTransp3}
\phi\frac{\partial\langle c\rangle^f}{\partial t}+\phi\langle\mathbf{u}\rangle^f\cdot\nabla\cdot\langle c\rangle^f=\phi\left(\underline{\mathcal{D}}+\underline{\boldsymbol\tau}+\underline{\mathbf{D}}\right):\nabla^2\langle c\rangle^f.
\end{equation}
Because $\nabla^2$ is a symmetric second-order tensor, only the symmetric part of the tensor in braces contributes to the transport process. We call this symmetric tensor the total dispersivity tensor
\begin{equation}\label{Eq:disp}
\underline{\mathbf{D}^*}=\frac{1}{2}\left(\underline{\mathcal{D}}+\underline{\boldsymbol\tau}+\underline{\mathbf{D}}\right)+\frac{1}{2}\left(\underline{\mathcal{D}}^T+\underline{\boldsymbol\tau}^T+\underline{\mathbf{D}}^T\right).
\end{equation}
We note that in the case where the image is fully resolved, $\varepsilon=1.0$, $D^*=D$, and the dispersivity tensor is identical to the one presented in \cite{2021-Soulaine}.

\section{Upscaling}
\label{section:upscaling}
In order to use effective dispersivity tensor in upscaled numerical simulation, it is often characterised as a function of the P\'eclet number $Pe$, which quantifies the balance between advective and diffusive transport. In this work, $Pe$ is calculated as,
\begin{equation}\label{Equ:Pe}
Pe=\frac{\langle\mathbf{u}\rangle^fL}{D},
\end{equation}
where $L$ is a characteristic length. In this work, we use the pore-scale length, defined as
\begin{equation}\label{equ:L}
L=\sqrt{\frac{\lambda K}{\phi}},
\end{equation}
where $\lambda$ is a constant equal 12 for 2D porous media and 8 for 3D porous media. This constant insures that, for homogeneous porous media made of capillary bundles, the pore-scale length is equal to the width of the pores. The permeability in the direction of a unit vector $\mathbf{e}_0$ can be calculated by solving the velocity field induced by an external body force equal to the mean pressure gradient,
\begin{equation}
\mathbf{S}_{uf}=\frac{\Delta P}{L_0}\mathbf{e}_0,
\end{equation}
where $\Delta P$ is the pressure drop and $L_0$ the length of the sample in the direction of $\mathbf{e}_0$. The flow regime is characterized by the Reynolds number $Re$,
\begin{equation}
Re=\frac{\langle\mathbf{u}\rangle^fL}{\nu}.
\end{equation}
If the flow is in the creeping flow regime, i.e. if $Re<<1$, then the permeability in the $\mathbf{e}_0$ direction can be calculated as
\begin{equation}
K_0=-\frac{\phi\langle\mathbf{u}\rangle^fL_0}{\nu\Delta P}.
\end{equation}

\section{Implementation}
\label{section:imp}
The numerical method is implemented in GeoChemFoam \cite{2022-Maes,2021c-Maes,2021-Maes}, our opensource reactive transport toolbox based on OpenFOAM\textsuperscript{\textregistered} \cite{OpenFOAM2016}. The full code can be downloaded from \href{www.github.com/geochemfoam}{www.github.com/geochemfoam}. The velocity field is calculated using \textit{simpleDBSFoam} \cite{2021c-Maes}. The standard solver \textit{scalarTransportFoam} has been modified into a new solver \textit{dispersionFoam} that solves Eq. (\ref{Eq:closureFinal}). The equation is discretized on a collocated Eulerian grid. The space discretization of the convection terms is performed using the first-order linear upwind scheme, while the diffusion term is discretized using the Gauss linear limited corrected scheme, which is second order and conservative. The closure problem is solved using the SIMPLE algorithm with under-relaxation with a coefficient of 0.7, and iterated until convergence with a residual lower than $10^{-6}$ \cite{OpenFOAM2016}. The dispersion tensor (Eq. (\ref{Eq:disp})) is calculated using a new post-processing utility called \textit{processDispersion}.

\section{Validation: multiscale micromodel}
\label{section:valid}

\begin{figure}[!t]
\begin{center}
\includegraphics[width=0.9\textwidth]{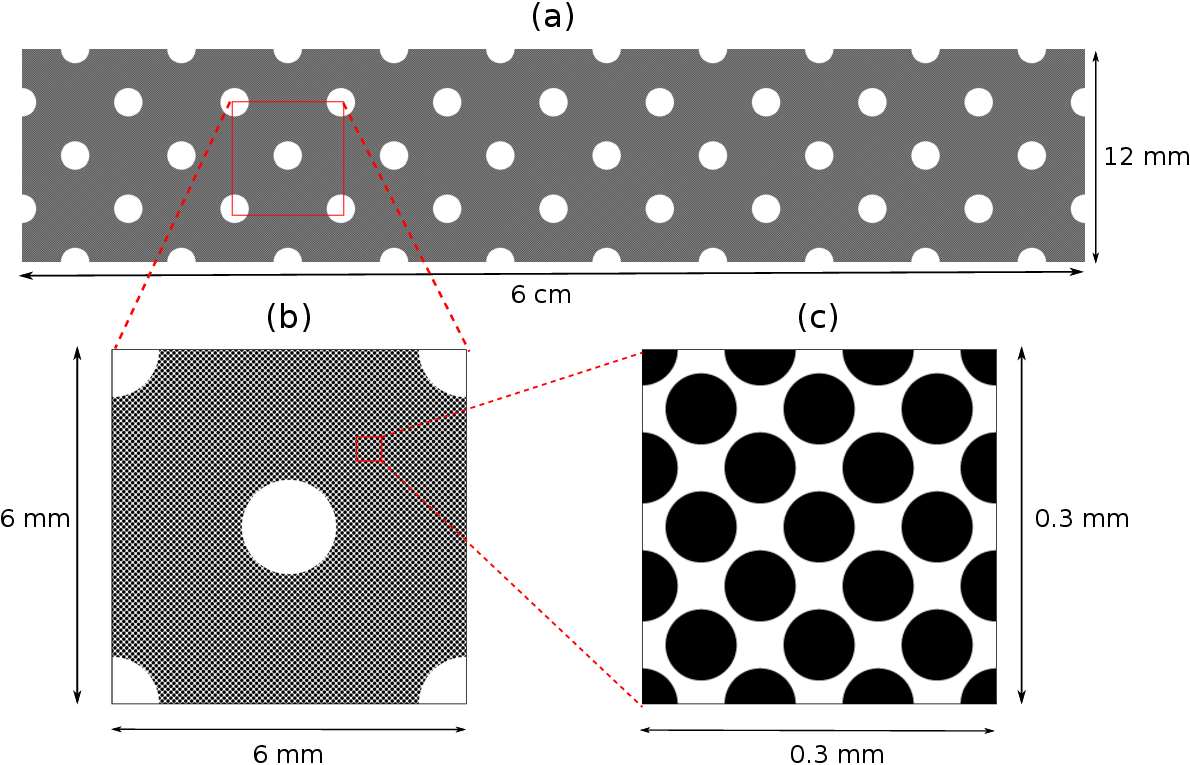}
\caption{Multiscale micromodel for numerical model validation. (a) Full micromodel. (b) Full model REV. (c) REV of matrix. \label{fig:TSmicromodel}}
\end{center}
\end{figure}

Our novel approach is validated by performing simulations on a 2D multiscale micromodel. The advantage of a 2D system is that we can build a high resolution image of a multiscale domain that can be an REV of every pore structure included while being small enough to perform fully resolved DNS simulation and compare with the micro-continuum approach and verify that our dispersivity calculation is correct. For this, we construct a 2D image representative of a vuggy carbonate rock, with homogeneous distribution of vugs and pores. The image is 1200$\times$6000 voxels with resolution 10 $\mu$m, and is presented in Fig. \ref{fig:TSmicromodel}a. A homogeneous system of large circular pores (i.e. vug) is added to a homogeneous matrix of smaller pores. The vugs are 1.6 mm in diameter and are organised into 5 lines and 21 columns. The distance between two lines and between two columns is 3 mm, but the vugs are offset, so that the distance between two successive vugs in the same line or the same column is 6 mm. Due to symmetry, the subdomain highlighted in a red square in Fig. \ref{fig:TSmicromodel}a is a REV of the full domain. A high resolution image of the REV is presented in Fig. \ref{fig:TSmicromodel}b. The matrix is made of circular beads with 60 $\mu$m diameters, organised in 121 lines and columns. Due to symmetry again, the subdomain highlighted in a red square in Fig. \ref{fig:TSmicromodel}b is a REV of the matrix (Fig. \ref{fig:TSmicromodel}c). The distance between two lines and between two columns is 50 $\mu$m, but since the beads are also offset, the distance between two successive beads in the same line or the same column is 100 $\mu$m. The pore bodies (i.e. void between two successive beads) have a maximum 40 $\mu$m inscribed diameter, and the pore throats (i.e. void between two offset beads) have a maximum 10.7 $\mu$m inscribed diameter. All data, images and base cases necessary to run the simulations can be downloaded from \href{www.github.com/geochemfoam/Examples}{www.github.com/geochemfoam/Examples}.

\begin{figure}[!t]
\begin{center}
\includegraphics[width=0.9\textwidth]{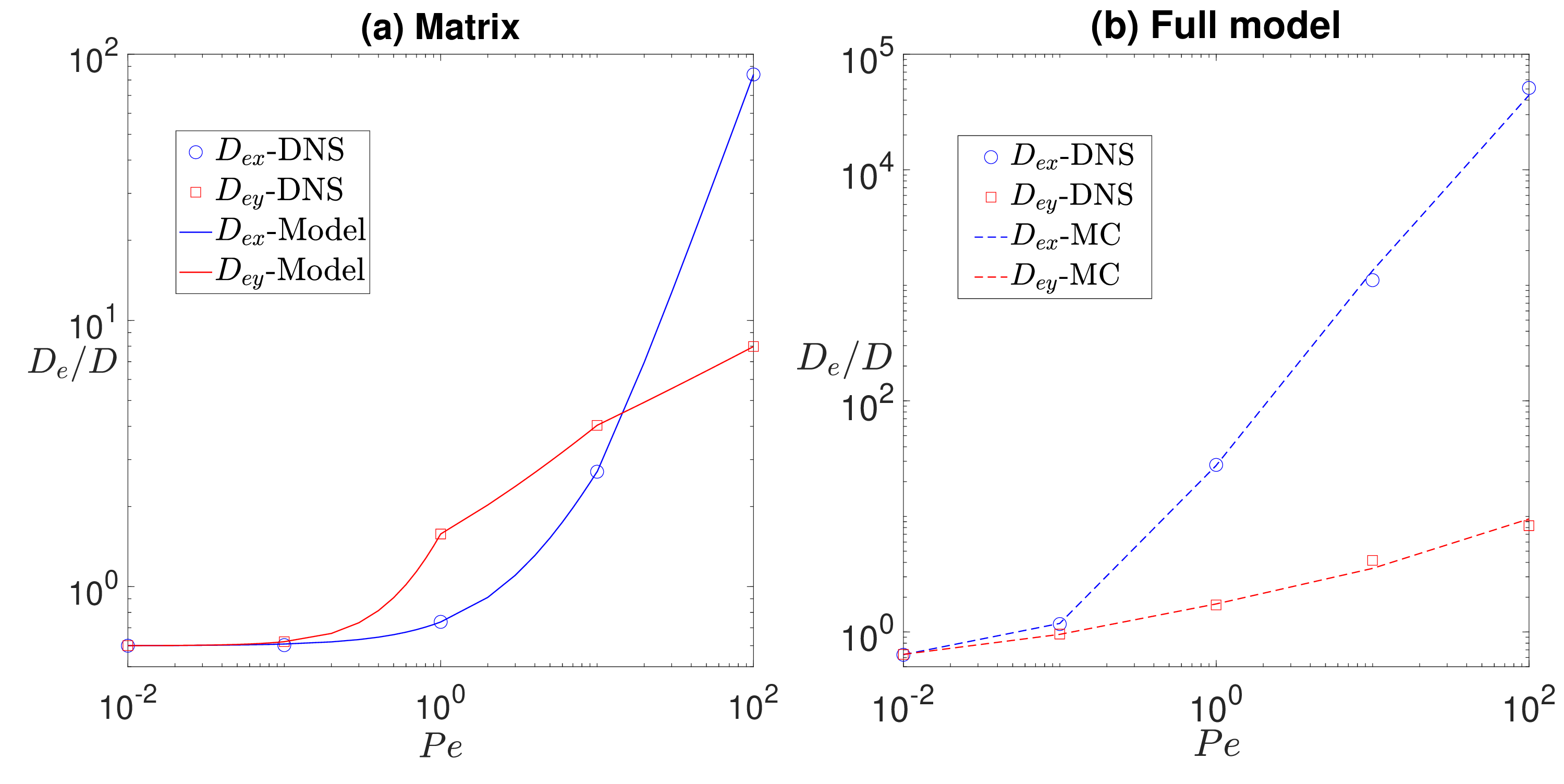}
\caption{Dispersivity in a multiscale 2D micromodel. (a) Dispersion coefficient in the matrix as a function of P\'eclet number obtained by high resolution DNS and fitted model (Eq. (\ref{Equ:dispModel})) and (b) Dispersion coefficient in the REV of the full model obtained by high resolution DNS and obtained using the micro-continuum approach.  \label{fig:dispmicromodel}}
\end{center}
\end{figure}

An effective dispersivity model for the matrix is obtained by performing DNS simulations on the REV (Fig. \ref{fig:TSmicromodel}c). The numerical model and simulation workflow are presented in \cite{2021-Soulaine}. The simulations are performed with mesh resolutions of 2 (150x150) and 1 $\mu$m(300x300) to verify mesh convergence. The domain is assumed to be cyclic between the left and right boundaries, and the top and bottom boundaries. All simulations in the matrix are run on 8 cores on an intel Xeon processor and take roughly 1 min to complete. Due to symmetry, the permeability and dispersivity tensors can be calculated with the flow in the x direction. The porosity and permeability of the porous matrix were numerically calculated as $\phi=0.429$ and $K=4.04\times10^{-12}$ m$^2$. The external body force (i.e. the mean pressure gradient) is tuned to obtain the velocity field at a Reynolds number Re=0.01. The longitudinal dispersion coefficient $D_{ex}$ and the transverse dispersion coefficient $D_{ey}$ are then calculated for $Pe=$0.01, 0.1, 1.0, 10 and 100 by tuning the diffusion coefficient following Eq. (\ref{Equ:Pe}). The results are shown in Fig. (\ref{fig:dispmicromodel}a).

In order to validate our micro-continuum approach, the dispersivity in the REV of the full model (Fig. \ref{fig:TSmicromodel}b) is calculated by two different methods. First, we use a high resolution grid for which all pores are resolved (incl. pores in the matrix) and the simulation model used previously (\cite{2021-Soulaine}) can be applied. The simulations are performed with mesh resolution of 2 (3000x3000) and 1 (6000x6000) $\mu$m to verify mesh convergence. The domain is assumed to be cyclic between the left and right boundaries, and the top and bottom boundaries. All simulations in the REV domain at high resolution are run on 2048 cores (16 nodes with 128 cores per node) on the UK national supercomputer ARCHER2. Like for the porous matrix, the permeability and dispersivity tensors can be calculated with the flow in the x direction due to symmetry. The velocity field is obtained by solving the Navier-Stokes equations in the fully-resolved pore-space and was completed in roughly 10 hours. The porosity and permeability were calculated numerically, obtaining $\phi=0.493$ and $K=5.10\times10^{-12}$ m$^2$, and the pressure gradient was tuned to obtain a Reynolds number Re=0.01. The longitudinal dispersion coefficient $D_{ex}$ and the transverse dispersion coefficient $D_{ey}$ are then calculated for $Pe=$0.01, 0.1, 1.0, 10 and 100 by tuning the diffusion coefficient following Eq. \ref{Equ:Pe}. Each simulation is run on 2048 cores and take roughly 50 min to complete.  The results are shown in Fig. (\ref{fig:dispmicromodel}b).

We then calculate dispersivity in the REV of the full model (Fig. \ref{fig:TSmicromodel}b) using our micro-continuum approach and compare to the results obtained from a fully resolved simulation. In order to integrate dispersivity in the porous matrix in our micro-continuum simulations, a model that calculates the dispersivity coefficients as a function of $Pe$ is developed. This model is obtained by fitting the best function of the form $D_e\propto 1+\beta Pe^\alpha$ for different intervals of $Pe$. The model is plotted on Fig. (\ref{fig:dispmicromodel}a) and defined as:
\begin{equation}\label{Equ:dispModel}
\begin{aligned}
& \text{if } Pe<1:& \\
& &D_{ex}=D\times 0.5995\left(1+0.228Pe^{1.1187}\right)\\
& &D_{ey}=D\times 0.5995\left(1+1.629Pe^{1.663}\right)\\
& \text{if } 1<Pe<10:& \\
& &D_{ex}=D\times 0.5995\left(1+0.228Pe^{1.1187}\right)\\
& &D_{ey}=D\times 0.5995\left(1+1.629Pe^{0.546}\right)\\
& \text{if } Pe>10:& \\
& &D_{ex}=D\times 0.5995\left(1+0.088Pe^{1.599}\right)\\
& &D_{ey}=D\times 0.5995\left(1+2.667Pe^{0.332}\right)\\
\end{aligned}
\end{equation}
The matrix dispersivity model is then integrated to Eq. (\ref{Eq:closureFinal}) and (\ref{Eq:disp}) in order to calculate dispersivity in the REV of the full model (Fig. \ref{fig:TSmicromodel}b) using our micro-continuum approach. We use a low-resolution grid of 10 $\mu$m (600x600). At this resolution, the pores in the porous matrix are not resolved. Instead, the porous matrix is represented by its porosity and permeability previously calculated ($\phi=0.429$ and $K=4.04\times10^{-12}$ m$^2$), and the dispersivity model presented in Eq. (\ref{Equ:dispModel}). All simulations in the REV model at low resolution are performed on 24 cores an intel Xeon processor and take roughly 2 minutes to complete. First, the velocity field is calculated by solving Eq. (\ref{Eq:continuityDBS}) and (\ref{Eq:momentumDBS}). We obtain the same porosity $\phi=0.493$ and permeability $K=5.10\times10^{-12}$ m$^2$ as for the high resolution model. The same conditions ($Re=0.01$, $Pe=$0.01, 0.1, 1.0, 10 and 100) are then reproduced. The closure variable is solved using Eq. (\ref{Eq:closureFinal}) and the longitudinal and transverse dispersion coefficients are calculated using Eq. (\ref{Eq:disp}). The coefficients calculated are presented in Fig. (\ref{fig:dispmicromodel}b). We observe that the coefficients calculated with the high resolution model and with the micro-continuum approach are almost identical.

\begin{figure}[!t]
\begin{center}
\includegraphics[width=0.9\textwidth]{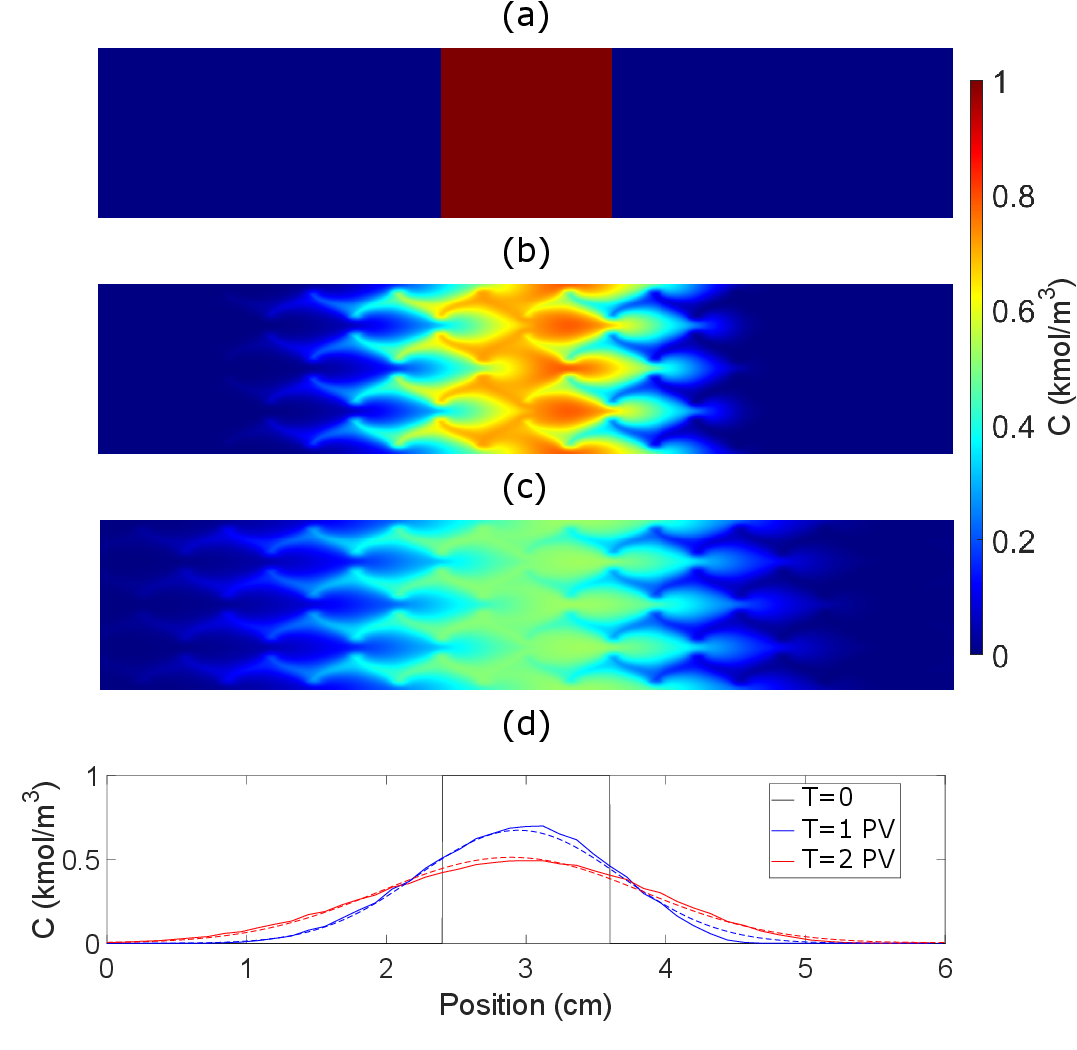}
\caption{Simulation of species transport in a 2D micromodel (Fig. \ref{fig:TSmicromodel}a) using the micro-continuum approach. (a) Initial concentration field. (b) Concentration field after one cycle (1 pore-volume injected). (c) Concentration field after two cycles (2 pore-volume injected). (d) Average concentration in the x-direction (plain lines) and concentration obtained with a one-dimensional upscaled model (dotted lines).\label{fig:longTSmicromodel}}
\end{center}
\end{figure}

To further verify that our calculation of dispersivity is accurate, we simulate the transport of species in the domain presented in Fig. \ref{fig:TSmicromodel}a. The species is initialized as presented in Fig. \ref{fig:longTSmicromodel}a, with a concentration equal 1 kmol/m$^3$ in a slug of length 1cm in the middle of the domain. The domain is assumed to be cyclic between the left and right boundaries, and the top and bottom boundaries. The conditions are defined such that $Re=0.01$ and $Pe=1$.

First, the velocity field and species transport are solved using the micro-continuum approach (Eq. (\ref{Eq:continuityDBS}), (\ref{Eq:momentumDBS}) and (\ref{Eq:concentrationDBS})) with a low-resolution grid of 20 $\mu$m (3000x600). The porous matrix is again represented by its porosity and permeability previously calculated ($\phi=0.429$ and $K=4.04\times10^{-12}$ m$^2$), and the dispersivity model presented in Eq. (\ref{Equ:dispModel}). The simulation is performed on 256 cores (2 nodes with 128 cores per nodes) on the ARCHER2 and takes roughly 10 minutes to complete. Fig. \ref{fig:longTSmicromodel}b and \ref{fig:longTSmicromodel}c shows the concentration field after 1 and 2 pore-volumes have been injected, respectively, i.e. after the slug has traveled one and two full cycles inside the domain. The average concentration along the x-direction is plotted in Fig. \ref{fig:longTSmicromodel}d (plain lines).

We then compare the results with the ones obtained using a homogeneous one-dimensional upscaled model with porosity $\phi=0.493$, permeability $K=5.10\times10^{-12}$ m$^2$ and dispersivity coefficients $D_{ex}=2.79\times10^{-7}$ m$^2$/s, which is the coefficient obtained with the closure problem for $Pe=1$. The concentration along the x-direction is plotted Fig. \ref{fig:longTSmicromodel}d (dotted lines). We observe that the one-dimensional model is able to match the evolution of the species concentration with good accuracy. 

We conclude that the micro-continuum approach can be employed to calculate dispersivity in multiscale porous materials using the closure problem presented in Eq. (\ref{Eq:closureFinal}) and (\ref{Eq:disp}). In the next part, the model will be applied to calculate dispersivity in digital twin of multiscale porous materials.

\section{Application: digital twin of multiscale materials}
\label{section:app}
In this section, our model is applied to calculate dispersivity in two different types of multiscale porous materials: hierarchical foams and microporous carbonate rocks. All data, images and base cases necessary to run the simulations can be downloaded from \href{www.github.com/geochemfoam/Examples}{www.github.com/geochemfoam/Examples}.
\subsection{Hierarchical foams}

\begin{figure}[!b]
\begin{center}
\includegraphics[width=0.95\textwidth]{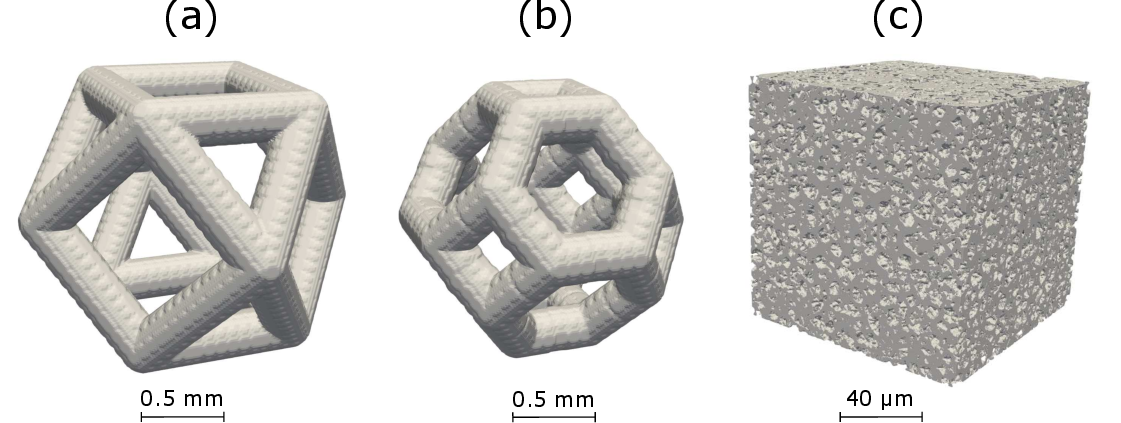}
\caption{3D images of Representative Elementary Volume for (a) Kelvin cell 1, (b) Kelvin cell 2 and (c) porous foam.\label{fig:KelvinCells}}
\end{center}
\end{figure}

Hierarchical foams can be manufactured by inscribing a porous foam onto a lattice, using templating processes or 3D printing \cite{2021-Kleger,2024-Mularczyk}. The emulsion within the foam generates a relatively homogenous and isotropic material inside the lattice, with pores typically ranging from 1 to 20 $\mu$m in diameters. The fibres in the lattice are separated by pores which are typically between 100 $\mu$m and 1 mm wide. Such materials are relevant for a number of applications such as catalysis, filtration, heat exchanger and energy storage devices due to their high porosity, high permeability, high surface area and high relative mechanical strength \cite{2016-Minas,2019-Huang}. Kelvin cell lattices are often employed due to their excellent mechanical properties \cite{2023-Karthic}. The efficiency of these processes requires that species penetrate inside the foam by diffusion or hydrodynamic dispersion.

To investigate materials of this type, we considered four scenario based on two different Kelvin cells (Kelvin cell 1 and Kelvin cell 2), and with either dense or porous fibers. REVs for two Kelvin cell unit lattices and for a porous foam were generated using the Porous Microstructure Generator \cite{PMG}. The Kelvin cells are 1 mm unit length and have fibers of uniform radius equal 100 $\mu$m. We use a resolution of 10 $\mu$m to ensure a good representation of the lattice geometry. For the porous fiber, we first generate a 300$^3$ voxel foam with a resolution of 0.2 $\mu$m, maximum particle radius of 4 $\mu$m and foam fibre radius of 0.6 $\mu$m. We then extract a 200$^3$ subvolume in the middle and use this to create a 400$^3$ cyclic domain by rotating and concatenating the lattice in all directions. The resulting domains are presented in Fig. \ref{fig:KelvinCells}.

\begin{figure}[!b]
\begin{center}
\includegraphics[width=0.95\textwidth]{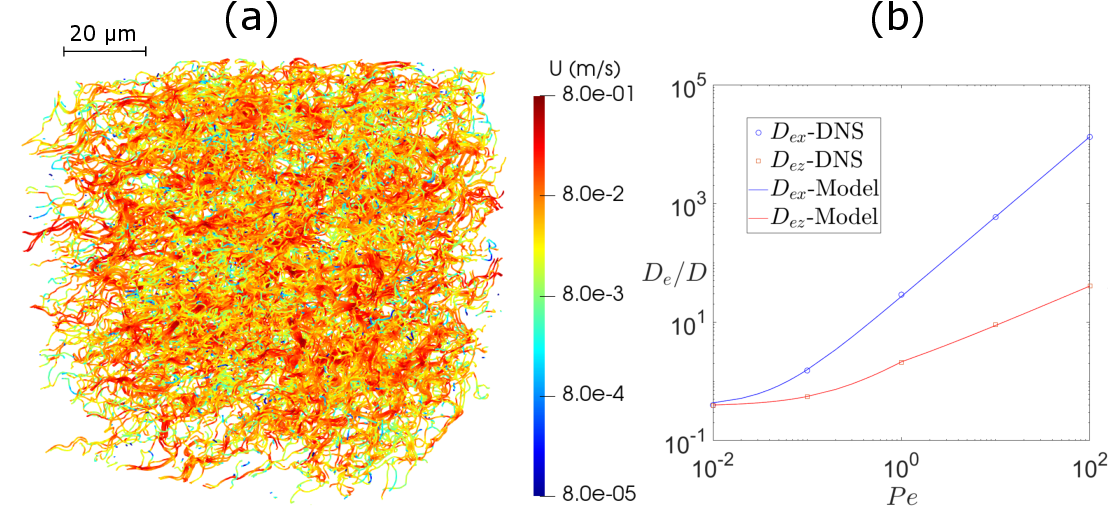}
\caption{Dispersivity in a porous foam (a) Streamlines obtained at a Reynolds number of 0.01 (b) Dispersion coefficient in the foam as a function of P\'eclet number obtained by DNS and numerical model obtained (Eq. (\ref{Equ:dispFoam})) \label{fig:FoamDisp}}
\end{center}
\end{figure}

An effective dispersivity model for the foam is obtained by performing DNS simulations on the domain. All simulations are run on 2048 cores (16 nodes with 128 cores per nodes) on ARCHER2 and take between 2 and 20 minutes to complete. Since the foam is mostly isotropic, the permeability and dispersivity tensors can be calculated with the flow in the x direction. The porosity and permeability are calculated as $\phi=0.445$ and $K=7.61\times10^{-15}$ m$^2$. The velocity field is established at a Reynolds number Re=0.01 and the streamlines are shown in Fig. \ref{fig:FoamDisp}a. The dispersion coefficients are then calculated for $Pe=$0.01, 0.1, 1.0, 10 and 100 and presented in Fig. \ref{fig:FoamDisp}b.

We then developed a model that calculates the dispersivity coefficients as a function of $Pe$ to be integrated in the micro-continuum approach by fitting the best function of the form $D_e\propto 1+\beta Pe^\alpha$ for different intervals of $Pe$. The model is plotted on Fig. (\ref{fig:FoamDisp}b) and defined as:
\begin{equation}\label{Equ:dispFoam}
\begin{aligned}
& \text{if } Pe<1:& \\
& &D_{ex}=D\times 0.38\left(1+70Pe^{1.35}\right)\\
& &D_{ey}=D_{ez}=D\times 0.38\left(1+4.6Pe^{1}\right)\\
& \text{if } Pe>1:& \\
& &D_{ex}=D\times 0.38\left(1+70Pe^{1.35}\right)\\
& &D_{ey}=D_{ez}=D\times 0.38\left(1+4.6Pe^{0.68}\right)\\
\end{aligned}
\end{equation}

\begin{figure}[!b]
\begin{center}
\includegraphics[width=0.95\textwidth]{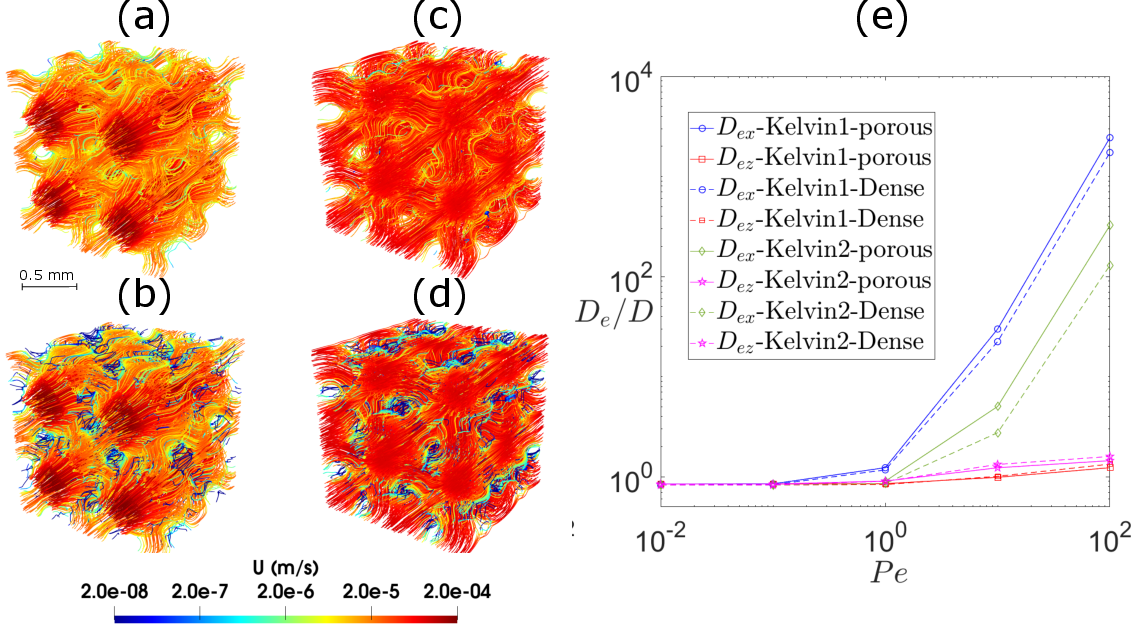}
\caption{Dispersion in hierarchical foams. Streamlines in (a) dense Kelvin cell 1; (b) porous Kelvin cell 1; (c) dense Kelvin cell 2; (d) porous Kelvin cell 2 at a Reynolds number of 0.01. (e) Calculated dispersion coefficient as a function of P\'eclet number for the four scenarios.\label{fig:KelvinDisp}}
\end{center}
\end{figure}

The numerical dispersion model for the foam can then be integrated to the model for the lattice using the micro-continuum approach.  All simulations are run on 256 cores (2 nodes with 128 cores per nodes) on ARCHER2 and take between 10 minutes and 1 hour to complete. Due to symmetry, the permeability and dispersivity tensors can be calculated with the flow in the x direction. The porosity and permeability are calculated and presented in Table \ref{tab:Kelvin}. In order to investigate dispersivity for the same pore velocity, the characteristic length scale is assumed to be equal to L=250 $\mu$m. The velocity field is established at a Reynolds number Re=0.01 and the streamlines are shown in Fig. \ref{fig:KelvinDisp}. The streamlines are similar between dense and porous cells, with the exception of small velocities in the porous fibers, four orders of magnitude smaller than the velocities in the lattice channels. The dispersion coefficients are then calculated for $Pe=$0.01, 0.1, 1.0, 10 and 100 and presented in Fig. \ref{fig:KelvinDisp}b.

\begin{table}[!t]
\centering
\begin{tabular}{c|c|c|c|c}
 & Kelvin 1 - dense & Kelvin 1 - porous & Kelvin 2 - dense & Kelvin 2 - porous\\[0.1cm]
\hline
Porosity & 0.78 & 0.88 & 0.78 & 0.88 \\[0.1cm]
Permeability ($\times 10^{-9}$m$^2$) & 6.6 & 6.9 & 5.1 & 5.3 \\[0.1cm]
Pore-scale length ($\mu$m)  & 260 & 250 & 230 & 220 \\[0.1cm]
\end{tabular}
\caption{Porosity and permeability of Kelvin cells used in this study.\label{tab:Kelvin}}
\end{table}

We observe that the dispersivity of the Kelvin cells 1 is always larger than the dispersivity of the Kelvin cells 2. This means that the spread of a species transported into a foam of Kelvin 1 structure will be greater than the spread for a foam of Kelvin 2 structure. However, in this study, we are more interested in the difference between the dispersivity of dense and porous structure, because it translates to species that invade the porous matrix, which will directly affect the system efficiency for many applications. Here, we notice that the difference between dispersivity with dense and porous matrix is greater for Kelvin 2 than for Kelvin 1. This suggests that more species will invade the porous matrix in the case of Kelvin 2.

\begin{figure}[!t]
\begin{center}
\includegraphics[width=0.95\textwidth]{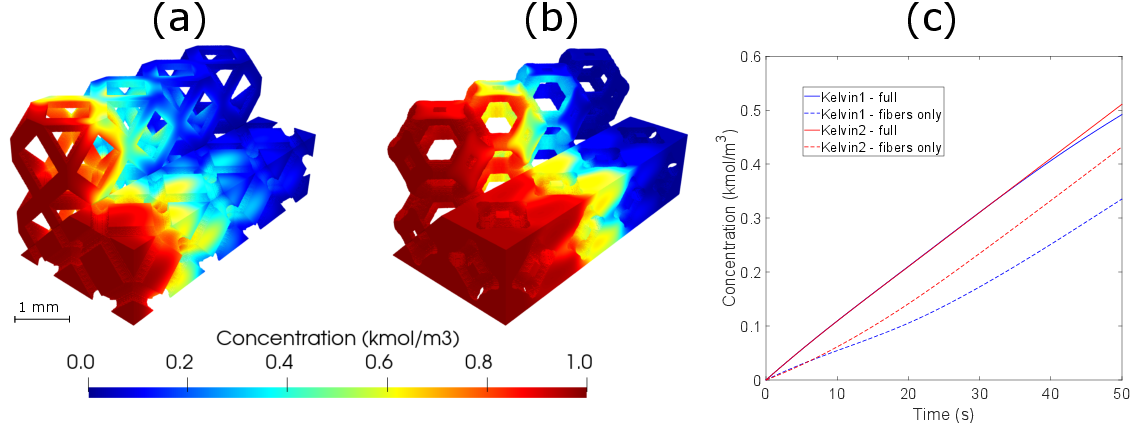}
\caption{Dispersive transport ($Pe=10$) in hierarchical foams. Concentration map in (a) porous Kelvin cell 1 and (b) porous Kelvin cell 2 after 50s (0.5 pore-volume injected). The top right corner of the foams has been removed to see the concentration inside. The lower diagonal shows only the cell in the lattice channels while the upper diagonal shows only the cells in the fibers; (c) Concentration as a function of time.\label{fig:KelvinConcentration}}
\end{center}
\end{figure}

To verify this, we perform a simulation of species transport in the porous Kelvin cells using the micro-continuum approach (Eq. (\ref{Eq:concentrationDBS}). The geometries are 4x2x2mm, i.e. 4x2x2 Kelvin cells. We use a resolution of 10 $\mu$m. We inject at a constant pressure drop that corresponds to a pore velocity of 4$\times 10^{-5}$ m/s. The molecular diffusion is set to $D=10^{-9}$ m$^2$/s, which corresponds to $Pe=10$. The species concentration is set to 1 at the inlet and a zero gradient boundary condition is applied at the outlet. Figures \ref{fig:KelvinConcentration}a and b show the concentration maps after 50s, i.e. when the total volume of species injected is equal to 0.5 the total pore-volume of the system. The top right corner of the foams has been removed to see the concentration inside. The lower diagonal shows only the cell in the lattice channels while the upper diagonal shows only the cells in the fibers. We observe that the concentration is more diffuse in the Kelvin cell 1 than in the Kelvin cell 2. Figure \ref{fig:KelvinConcentration}c shows the time evolution of the total concentration (plain lines) in the domain as well as the concentration in the fiber (dashed line). We observe that up to T=40s, the total concentration in the domain is almost identical. This is because the species has not reached the right boundary of the system. At T=50s, the species has reached the boundary for the Kelvin cell 1 but not for the Kelvin cell 2. This is because of the higher dispersion in the Kelvin cell 1, as observed on Figures \ref{fig:KelvinConcentration}a and b. The species front is more diffuse for Kelvin cell 1 and sharper for Kelvin cell 2. We observe that the concentration in the fibers is larger for Kelvin cell 2 than for Kelvin cell 1, even before the species reach the outlet. This significantly higher concentration comes from the difference between the dispersivity with and without porous fibers, which is greater for Kelvin cell 2.

Our investigation suggests that, for reactive transport processes such as adsorption filters and columns, a porous foam of Kelvin cell 2 structure will be more efficient than one with Kelvin cell 1 structure, since more species will be transported inside the foam where it can react. The flow and transport properties calculated in this study can be used in continuum-scale modelling to investigate process performance \cite{2015-Soares}. Other factors, such as mechanical strength, need to be considered to fully optimise the system, but this is out of the scope of this study.

\subsection{Microporous carbonate rock}

Carbonate subsurface formations, such as depleted oil and gas field, are being considered for carbon storage, due to their prevalence, in particular in the Middle East and the North Sea. These porous materials are some of the most complex involved in engineering applications, where several pore structures such as nanopores, micropores, macropores and fractures form systems that are often only connected when described together. Dispersivity is of particular interest for carbon storage as it strongly impacts reactive transport with precipitation and dissolution that may have strong benefits (e.g., CO$_2$ mineralization, permeability enhancement) or generate high engineering risks (e.g., pore clogging, reservoir integrity).

Carbonate rocks are often multiscale and therefore a multiscale representation is necessary to estimate their properties. Menke et al. \cite{2022-Menke} uses a multiscale description that includes 14 phases (e.g., pores, solid and 12 microporous phases) to estimate the flow properties of a sample of the Estaillades carbonate formation and compare with experiments. The micro-CT image was of size 1202x1236x6000 voxels wtih resolution 3.9676 microns. In order to estimate the permeability of each of the 12 microporous phases, a nano-CT image with resolution 32 nanometers was obtained from a laser-cut sub-sample. The information from the image was then used as input into an object-based pore network generator, on which permeability fields were simulated for a range of porosities, creating a synthetic porosity–permeability relationship that was used for each phase.

\begin{figure}[!b]
\begin{center}
\includegraphics[width=0.95\textwidth]{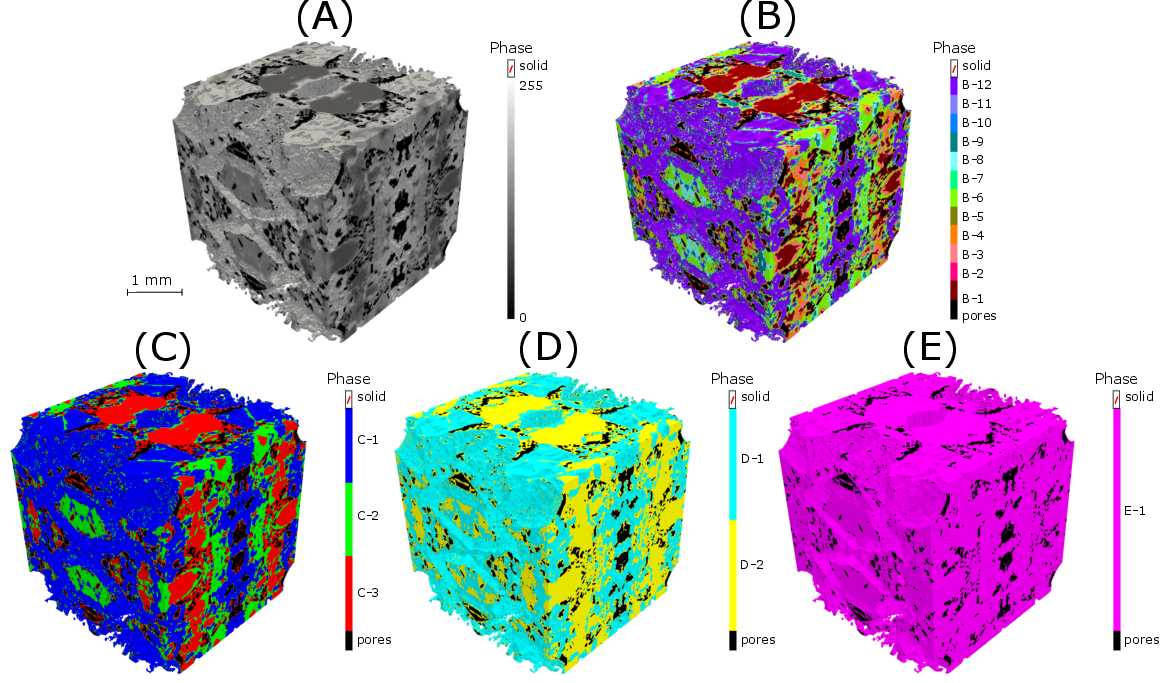}
\caption{Estaillades micro-CT cyclic subvolume (600x600x600 voxels); (A) Greyscale; (B) 14 phases segmentation; (C) 5 phases segmentation; (D) 4 phases segmentation; (E) 3 phases segmentation; The solid phase has been made transparent. The properties of the phases are presented in Table \ref{Tab:EstailladesMicro}. \label{fig:EstailladesMacroPhases}}
\end{center}
\end{figure} 

In this work, we investigate the minimum number of phases that are necessary to include in the system to accurately calculate porosity, permeability, and dispersivity. For this we consider a subvolume of the Estaillades micro-CT image from \cite{2022-Menke} with size $300^3$ voxels extracted from the middle. The image is then rotated and combined in all directions to obtain cyclic $600^3$ image. The greyscale and segmented (into 14 phases) images are presented in Fig. \ref{fig:EstailladesMacroPhases}A and \ref{fig:EstailladesMacroPhases}B. The 12 microporous phases are labelled from B-1 to B-12. Their volume fraction and porosity are presented in Table \ref{Tab:EstailladesMicro}.

In order to investigate the possibility of simplifying the system, the microporous phases are regrouped into 3 (Fig. \ref{fig:EstailladesMacroPhases}C), 2 (Fig. \ref{fig:EstailladesMacroPhases}D) and 1 (Fig. \ref{fig:EstailladesMacroPhases}E) phases. The porosity of the new phases are calculated in order to match the average porosity of the group, and are presented in Table \ref{Tab:EstailladesMicro}.

\begin{table}[!b]
\centering
\begin{tabular}{c|c|c|c|c|c|c|c|c|c|c}
 label & Vol. frac. & Group & Porosity & Perm. (m$^2$) & $\tau$ & $\alpha$ & $\beta_1$ & $\gamma_1$ & $\gamma_2$ & $\beta_2$\\[0.15cm]
\hline
pores & 0.100 & NA & 1.0 & NA & NA & NA & NA & NA & NA & NA \\[0.1cm]
B-1 & 0.140 & NA & 0.57 & 7.3e-15 & 1.56 & 1.68 & 10 & 1.26 & 0.50 & 0.84 \\[0.1cm]
B-2 & 0.044 & NA & 0.52 & 5.3e-15 & 1.67 & 1.62 & 15 & 1.18 & 0.52 & 0.98 \\[0.1cm]
B-3 & 0.047 & NA & 0.47 & 4.0e-15 & 1.79 & 1.60 & 22 & 1.16 & 0.54 & 1.26 \\[0.1cm]
B-4 & 0.052 & NA & 0.42 & 3.2e-15 & 1.96 & 1.58 & 30 & 1.10 & 0.56 & 1.48 \\[0.1cm]
B-5 & 0.085 & NA & 0.36 & 1.8e-15 & 2.08 & 1.56 & 71 & 1.00 & 0.58 & 2.00 \\[0.1cm]
B-6 & 0.089 & NA & 0.27 & 7.4e-16 & 2.44 & 1.54 & 250 & 0.98 & 0.60 & 3.40 \\[0.1cm]
B-7 & 0.041 & NA & 0.22 & 4.1e-16 & 2.78 & 1.54 & 570 & 0.95 & 0.62 & 4.10 \\[0.1cm]
B-8 & 0.032 & NA & 0.18 & 2.4e-16 & 3.22 & 1.58 & 1300 & 0.92 & 0.64 & 5.20 \\[0.1cm]
B-9 & 0.027 & NA & 0.15 & 1.4e-16 & 3.70 & 1.62 & 3000 & 0.94 & 0.66 & 7.20 \\[0.1cm]
B-10 & 0.024 & NA & 0.12 & 7.2e-17 & 4.55 & 1.66 & 9000 & 0.98 & 0.68 & 10.9 \\[0.1cm]
B-11 & 0.021 & NA & 0.09 & 2.7e-17 & 5.56 & 1.68 & 4.1$\times 10^4$ & 1.04 & 0.72 & 15.9 \\[0.1cm]
B-12 & 0.076 & NA & 0.07 & 1.1e-17 & 6.67 & 1.70 & 1.7$\times 10^5$ & 1.08 & 0.77 & 16.6 \\[0.1cm]
\hline
C-1 & 0.282 & B-1 to B-4 & 0.52 & 5.3e-15 & 1.67 & 1.62 & 15 & 1.18 & 0.52 & 0.98 \\[0.1cm]
C-2 & 0.247 & B-5 to B-8 & 0.28 & 8.1e-16 & 2.38 & 1.54 & 210 & 1.0 & 0.60 & 3.20 \\[0.1cm]
C-3 & 0.147 & B-9 to B-12 & 0.10 & 3.4e-17 & 5.26 & 1.67 & 4.1$\times 10^4$ & 0.99 & 0.59 & 2.56 \\[0.1cm]
\hline
D-1 & 0.456 & B-1 to B-6 & 0.44 & 3.6e-15 & 1.85 & 1.59 & 28 & 1.14 & 0.54 & 1.40 \\[0.1cm]
D-2 & 0.220 & B-7 to B-12 & 0.13 & 7.7e-16 & 4.00 & 1.65 & 5200 & 0.96 & 0.67 & 8.50 \\[0.1cm]
\hline
E-1 & 0.676 & B-1 to B-12 & 0.34 & 1.5e-15 & 2.22 & 1.55 & 100 & 1.26 & 0.5 & 0.84 \\[0.1cm]
\end{tabular}
\caption{Properties of segmentation phases for Estaillades. \label{Tab:EstailladesMicro}}
\end{table}

In order to obtained a representative image for each phase, the nano-CT image from Menke et al. \cite{2022-Menke} is eroded by applying a velocity to the fluid/solid interface, constant in magnitude and parallel to the interface. The porosity is then increased or decreased (depending on the sign of the velocity) until it matches the porosity of the phase presented in Table \ref{Tab:EstailladesMicro}. The method is implemented in an OpenFOAM utility \textit{erodeFoam} (\href{www.github.com/geochemfoam}{www.github.com/geochemfoam}). We then extract a $300^3$ image from the middle. The image is then rotated and combined in all directions to obtain a cyclic $600^3$ image. We assume that the images obtained are REVs of all phases. The images are then used to calculate the flow and transport properties of each phase. The same method that was applied in the previous subsection is employed, and the permeability, and dispersion coefficients for Pe=0.01, 0.1, 1.0, 10 and 100, are calculated. All simulations are performed with uniform grid with 216M cells, and are run on 2048 cores (16 nodes with 128 cores per nodes) on ARCHER2 and take between 1 and 6 hours to run. The dispersion coefficients are fitted with the following model:
\begin{equation}\label{Equ:dispModelEst}
\begin{aligned}
& \text{if } Pe<1:& \\
& &D_{ex}=\frac{D}{\tau}\left(1+\beta_1 Pe^{\alpha_1}\right)\\
& &D_{ey}=D_{ez}=\frac{D}{\tau}\left(1+\beta_2Pe^{\gamma_1}\right)\\
& \text{if } Pe>1:& \\
& &D_{ex}=\frac{D}{\tau}\left(1+\beta_1 Pe^{\alpha_2}\right)\\
& &D_{ey}=D_{ez}=\frac{D}{\tau}\left(1+\beta_2Pe^{\gamma_2}\right)\\
\end{aligned}
\end{equation}

\begin{figure}[!t]
\begin{center}
\includegraphics[width=0.95\textwidth]{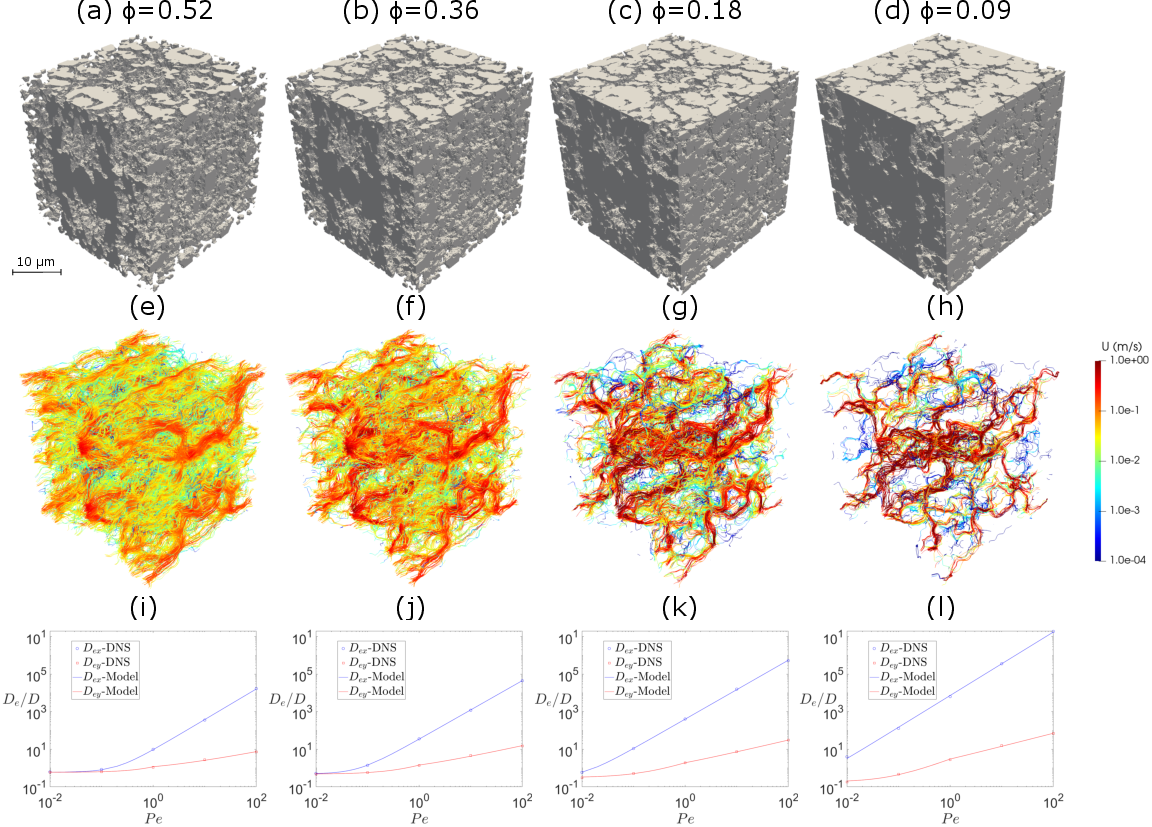}
\caption{Flow and transport properties for Estaillades microporous phases; 4 phases from Table \ref{Tab:EstailladesMicro} are presented (B-2, B-5, B-8 and B-11); (a) to (d) show the solid surfaces; (e) to (h) show the velocity field at Re=0.01; (i) to (l) show the dispersion coefficients calculated using DNS and fitted with the model from Eq. (\ref{Equ:dispModelEst})\label{fig:EstailladesMicro}}
\end{center}
\end{figure}

Fig. \ref{fig:EstailladesMicro} shows four of these images with their velocity field and dispersion coefficient as a function of the P\'eclet number. All properties are summarized in Table \ref{Tab:EstailladesMicro}.

\begin{figure}[!t]
\begin{center}
\includegraphics[width=0.95\textwidth]{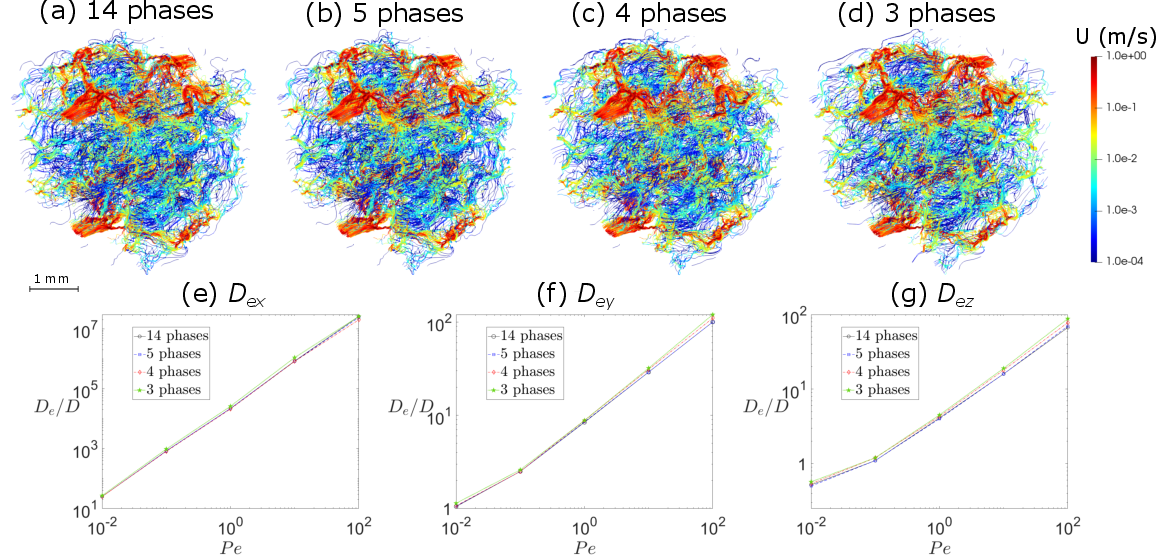}
\caption{Flow and transport properties for Estaillades micro-CT image; 4 segmentations are presented (Fig. \ref{fig:EstailladesMacroPhases}). (a) to (d) show the velocity field at Re=0.01;(e) to (g) shows the dispersion coefficients.\label{fig:EstailladesMacroU}}
\end{center}
\end{figure}

The properties are then used to calculate permeability and dispersion coefficients for the micro-CT image of Estaillades and the results obtained when using 14, 5, 4 and 3 segmentation phases are compared. Fig. \ref{fig:EstailladesMacroU} shows the velocity field (Re=0.01) and dispersion coefficients obtained for all segmentations. All simulations are performed with uniform grid with 216M cells, and are run on 2048 cores (16 nodes with 128 cores per nodes) on ARCHER2 and take between 2 and 12 hours to run. The results are summarized in Table \ref{Tab:EstailladesMacro}.

\begin{table}[!t]
\centering
\begin{tabular}{c||c|c|c||c|c|c||c|c|c||c|c|c}
& \multicolumn{3}{c||}{14 phases} & \multicolumn{3}{c||}{5 phases} & \multicolumn{3}{c||}{4 phases} & \multicolumn{3}{c}{3 phases}\\[0.15cm]
\hline
Poro.       & \multicolumn{3}{c||}{0.33} & \multicolumn{3}{c||}{0.33} & \multicolumn{3}{c||}{0.33} & \multicolumn{3}{c}{0.33} \\[0.1cm]
Perm.      & \multicolumn{3}{c||}{5.4} & \multicolumn{3}{c||}{5.3} & \multicolumn{3}{c||}{5.2} & \multicolumn{3}{c}{4.8} \\
 ($10^{-14}$ m$^2$)  & \multicolumn{3}{c||}{} & \multicolumn{3}{c||}{} & \multicolumn{3}{c||}{} & \multicolumn{3}{c}{} \\
\hline
$D_e$/$D$ & x & y & z & x & y & z & x & y & z & x & y & z \\
 Pe=0.01 & 25 & 1.0 & 0.52 & 25 & 1.0 & 0.53 & 25 & 1.1 & 0.54 & 27 & 1.2 & 0.57 \\
 Pe=0.1 & 840 & 2.5 & 1.1 & 840 & 2.5 & 1.1 & 810 & 2.5 & 1.2 & 970 & 2.6 & 1.2 \\
 Pe=1.0 & 2.2$\times 10^{4}$ & 8.4 & 4.1 & 2.2$\times 10^{4}$ & 8.4 & 4.0 & 2.1$\times 10^{4}$ & 8.8 & 4.3 & 2.6$\times 10^{4}$ & 8.8 & 4.5 \\
 Pe=10 & 8.4$\times 10^{5}$ & 29 & 16 & 8.4$\times 10^{5}$ & 30 & 16 & 8.2$\times 10^{5}$ & 31 & 18 & 1.1$\times 10^{6}$ & 32 & 19 \\
 Pe=100 & 2.6$\times 10^{7}$ & 100 & 67 & 2.5$\times 10^{7}$ & 110 & 70 & 2.0$\times 10^{7}$ & 110 & 78 & 2.6$\times 10^{7}$ & 120 & 88
 \end{tabular}
\caption{Properties of Estaillades micro-CT for various segmentation. \label{Tab:EstailladesMacro}}
\end{table}

As observed in Fig. \ref{fig:EstailladesMacroU}, the velocity field is not strongly impacted by the number of segmentation phases. This suggest that the flow structure is mostly impacted by the distribution between pores, micropores and solid, as represented when using a 3 phase segmentation. Some complexity in the velocity field within the micropores is not accurately represented for the 3 phase segmentation, leading to a slightly lower permeability (4.8 $\times 10^{-14}$ m$^2$ instead of 5.4  $\times 10^{-14}$ m$^2$).

Similarly, the dispersion coefficients calculated using segmentation with 3 phases, 4 phases and 5 phases are close to the one calculated with 14 phases. For 3 phases, the coefficients are slightly higher. For 4 phases, $D_{ex}$ is slightly lower while $D_{ey}$ and $D_{ez}$ are slightly higher. The coefficients obtained with 5 and 14 phases are almost identical. These coefficients can then be used in core-scale modelling based on Darcy's law \cite{2021-Wenck}.

We conclude that 12 microporous phases are not necessary to capture the flow and transport properties. A segmentation that includes 3 microporous phases is enough to calculate accurately permeability and dispersivity coefficients for $Pe$ from 0.01 to 100. In this case, even a segmentation with only one microporous phase is enough to calculate the coefficients with only a small margin of error. However, this might change for a different rock, or for multiphase flow.

\section{Conclusion}

In this work, we presented a novel numerical model, based on the micro-continuum approach, to compute dispersivity in multiscale porous materials. The equations based on the closure method were presented and validated by comparison with high resolution DNS for a 2D micromodel with multiscale porosity. The model was then applied to calculate dispersivity in two different types of geometry: hierarchical foam and microporous carbonate rock. For the hierarchical foam, we show that a Kelvin 2 lattice structure was more efficient than a Kelvin 1 to disperse species into the foam, which could lead to more adsorption, for example for filtration or catalyst. For the microporous rock, we demonstrated that only 3 microporous phase were necessary to calculate accurately flow and transport properties, a significant drop compared to the 12 microporous phase in the original model.

Although the method presented is accurate and generic, it does rely on the development of dispersion models for the under-resolved structure. The models used in our work are based on several assumptions: (1) the under-resolved porosity is isotropic, (2) the dispersion coefficients only depends on the velocity in the direction of the average flow and (3) the dispersion coefficients can be fitted in a model that only depends on the P\'eclet number. These assumptions are somewhat more accurate in simple geometries (e.g. hierarchical foams) than for complex systems (e.g., microporous carbonate rocks). The advantage of our method, however, is that a large number of simulations with varying properties such as porosity, velocity magnitude and direction can be performed to characterise the dispersion coefficients in the under-resolved phases and then integrated in our machine-learning based upscaling workflow \cite{2021-Menke}, which would be a target of future work.

\section*{Acknowledgment}
This work was funded under the EPSRC ECO-AI grant (references: EP/Y006143/1, EP/Y005732/1) and the NERC GeoSafe grant (reference NE/Y002504/1). The authors thank Prof. John Brodholt and the ARCHER2 Mineral and Geophysics consortium for providing the necessary computing credits, and Dr Cyprien Soulaine for sharing his DNS code, as well as for the valuable insight on how the micro-continuum method works.

\bibliographystyle{spphys} 
\bibliography{mybibliography.bib}


%\begin{thebibliography}{00}



%\end{thebibliography}
\end{document}